\documentclass{elsart}
\usepackage{epsfig}
\usepackage{enumerate}
\usepackage{amsmath}

\begin{document}
\begin{frontmatter}

\title{Phase transitions and universality in the Sznajd model with anticonformity}

\author{Matheus Calvelli $^{1,2}$,}
\thanks{matheuscalvelli@id.uff.br}
\author{Nuno Crokidakis $^{2}$}
\thanks{nuno@if.uff.br}
\author{and Thadeu J. P. Penna $^{1}$}
\thanks{thadeupenna@id.uff.br}

\address{
$^{1}$ICEx, Universidade Federal Fluminense, Volta Redonda/RJ, Brazil \\  
$^{2}$Instituto de F\'{\i}sica, Universidade Federal Fluminense, Niter\'oi/RJ, Brazil
}

\maketitle

\begin{abstract}
\noindent
In this work we study the dynamics of opinion formation in the Sznajd model with anticonformity on regular lattices in two and three dimensions. The anticonformity behavior is similar to the introduction of Galam's contrarians in the population. The model was previously studied in fully-connected networks, and it was found an order-disorder transition with the order parameter exponent $\beta=1/2$ calculated analytically. However, the other phase transition exponents were not estimated, and no discussion about the possible universality of the phase transition was done. Our target in this work is to estimate numerically the other exponents $\gamma$ and $\nu$ for the fully-connected case, as well as the three exponents for the model defined in square and cubic lattices. Our results suggest that the model belongs to the Ising model universality class in the respective dimensions.

\end{abstract}

\end{frontmatter}


\section{Introduction}

Models of opinion formation have been studied by physicists since the 80's and are now part of the new branch of physics called sociophysics. This recent research area uses tools and concepts of  statistical physics to describe some aspects of social and political behavior \cite{galam_book,sen_book,loreto_rmp}. From theoretical point of view, opinion models are interesting to physicists because they present order-disorder transitions, scaling and universality, among other typical features of physical systems, which called the attention of many groups throughout the world \cite{lalama,galam_rev,schneider,lccc,andre,javarone1,javarone2,xiong,sibona,xiong1,xiong2,javarone3}.

The standard approach to many models of opinion dynamics is to consider that the individuals are susceptibles, i.e., they can change opinion due to interactions with other individuals in the population \cite{galam_book,sen_book}. In such case, it is usual to define a network of contacts among individuals and the rules of microscopic interaction. The most famous models considering opinions as discrete or continuous variables are the Sznajd model \cite{sznajd_original}, the Galam model \cite{galam_1986,galam_1999}, the CODA model \cite{andre}, the kinetic exchange opinion models \cite{lccc,biswas}, the Deffuant model \cite{deffuant}, the Hegselmann-Krause model \cite{hk}, among others. Susceptible agents are also known as conformist individuals. On the other hand, there are also individuals that act in an opposite way in comparison with the conformist, called anticonformists. Anticonformists are similar to conformists, since both take cognizance of the group norm. Whereas conformists agree with the norm, anticonformers disagree. Galam created the term ``contrarians'' to refer to those individuals \cite{galam_cont}. The presence of such contrarians/anticonformists usually lead the model in consideration to undergoe order-disorder phase transitions \cite{galam_book,sen_book,loreto_rmp,sznajd_anti,maj_vote_anti}.

In this work we study the effects of anticonformity in the dynamics of the Sznajd model. The model is defined on square and cubic lattices, and also in the fully-connected newtork. Our motivation is to explore in details the critical behavior of the model, since the original work \cite{sznajd_anti} only considered the mean-field formulation of the model, and it did not explored the critical behavior of the model. We estimate the critical exponents numerically for the mentioned lattices, and we verified that the model belongs to the universality class of the Ising model on the same lattices.

This work is organized as follows. In Section 2 we present the microscopic rules that define the Sznajd model with anticonformity, and our numerical results on regular lattices. Finally, our conclusions are presented in section 3.


\section{Model}

The Sznajd model with anticonformity was studied recently. The authors defined it in the fully-connected graph \cite{sznajd_anti}. Let us consider a set of $N$ individuals (or agents), each one described by an Ising-like variable $o_{i}=+1 \,(\uparrow)$ or $o_{i}=-1 \,(\downarrow)$ ($i=1,2,...,N$), denoting two opposite opinions, for example the vote for two distinct candidates A or B. The initial concentration of each opinion is $0.5$ (disordered state). At each time step two neighbor agents (for example $i$ and $j$) are chosen, and they influence a third neighbor (for example $k$) in the following way:

\begin{itemize}
\item $\uparrow \uparrow \Downarrow \rightarrow \uparrow \uparrow \Uparrow$, conformity with probability $p_1$
\item $\downarrow \downarrow \Uparrow \rightarrow \downarrow \downarrow \Downarrow$, conformity with probability $p_1$
\item $\uparrow \uparrow \Uparrow \rightarrow \uparrow \uparrow \Downarrow$, anticonformity with probability $p_2$
\item $\downarrow \downarrow \Downarrow \rightarrow \downarrow \downarrow \Uparrow$, anticonformity with probability $p_2$
\end{itemize}

First two processes correspond to conformity, and the next two describe anticonformity. The authors in \cite{sznajd_anti} defined the ratio $r=p_{2}/p_{1}$ as the relevant parameter to control the phase transition of the model. Furthermore, the authors decided to investigate the case in which $p_{1}=1$ and $p_{2} \in [0,1]$ is the only parameter of the model. In this case, the control parameter is indeed $p_{2}$, since $r=p_{2}$ for $p_{1}=1$. As discussed in the original work \cite{sznajd_anti}, it can be easily shown that contrarian behavior introduced by Galam \cite{galam_cont} requires that $p_{1}+ p_{2} = 1$ and therefore is a special case of anticonformity.

In order to analyze the phase transition, we usually define the public opinion as the magnetization per site of the system, 
\begin{equation} \label{eq1}
m = \left\langle \frac{1}{N}\left|\sum_{i=1}^{N} o_{i}\right|\right\rangle ~,
\end{equation}  
\noindent
where $\langle\, ...\, \rangle$ denotes a disorder or configurational average, taken after a large enough number of whole-lattice sweeps. 

Our objective in this work is to analyze the critical behavior of the system. In addition to the above-mentioned order parameter $m$, Eq. \eqref{eq1}, we will also consider the susceptibily $\chi$ and the Binder cumulant $U$,  that were not considered in the previous work \cite{sznajd_anti}. Those quantities are defined respectively as 
\begin{eqnarray} \label{eq2}
\chi & = & N\,(\langle m^{2}\rangle - \langle m \rangle^{2}) ~, \\ \label{eq3}
U & = & 1 - \frac{\langle m^{4}\rangle}{3\,\langle m^{2}\rangle^{2}} \,.  
\end{eqnarray}

In order to estimate the critical exponents of the phase transition, we performed numerical simulations of the model for distinct population sizes $N$, and we considered a finite-size scaling (FSS) analysis of the numerical data, based on the standard FSS relations \cite{cardy}
\begin{eqnarray} \label{eq4}
O(N) & \sim & N^{-\beta/\nu} \\  \label{eq5}
\chi(N) & \sim & N^{\gamma/\nu} \\   \label{eq6}
U(N) & \sim & {\rm constant} \\   \label{eq7}
p_{2,c}(N) - p_{2,c} & \sim & N^{-1/\nu} ~,
\end{eqnarray}
that are valid in the vicinity of the transition. Based on Eqs. \eqref{eq4}-\eqref{eq7}, we can estimate the critical exponents $\beta$, $\gamma$ and $\nu$. In this case we will consider separately in the following subsections three distinct cases, namely the model defined on: (i) the fully-connected graph, (ii) the two-dimensional square lattice, and (iii) the three-dimensional cubic lattice.


\subsection{Fully-connected graph}

In this case, we consider that each agent can interact with all others, i.e., all the three indivuals $(i,j,k)$ are randomly chosen in the population. The analytical calculations performed in \cite{sznajd_anti} for the steady states of the model predicted that there is an order-disorder transition at a critical point $p_{2,c}=1/3$, and the behavior of the order parameter near that critical point is given by
\begin{equation} \label{eq8}
m = \sqrt{\frac{1-3p_{2}}{1+p_{2}}} ~, 
\end{equation}
\noindent
that can be rewriten in the form $m\sim (p_{2}-p_{2,c})^{\beta}$, with the typical Ising mean-field exponent $\beta=1/2$.

\begin{figure}[t]
\begin{center}
\vspace{3mm}
\includegraphics[width=0.48\textwidth,angle=0]{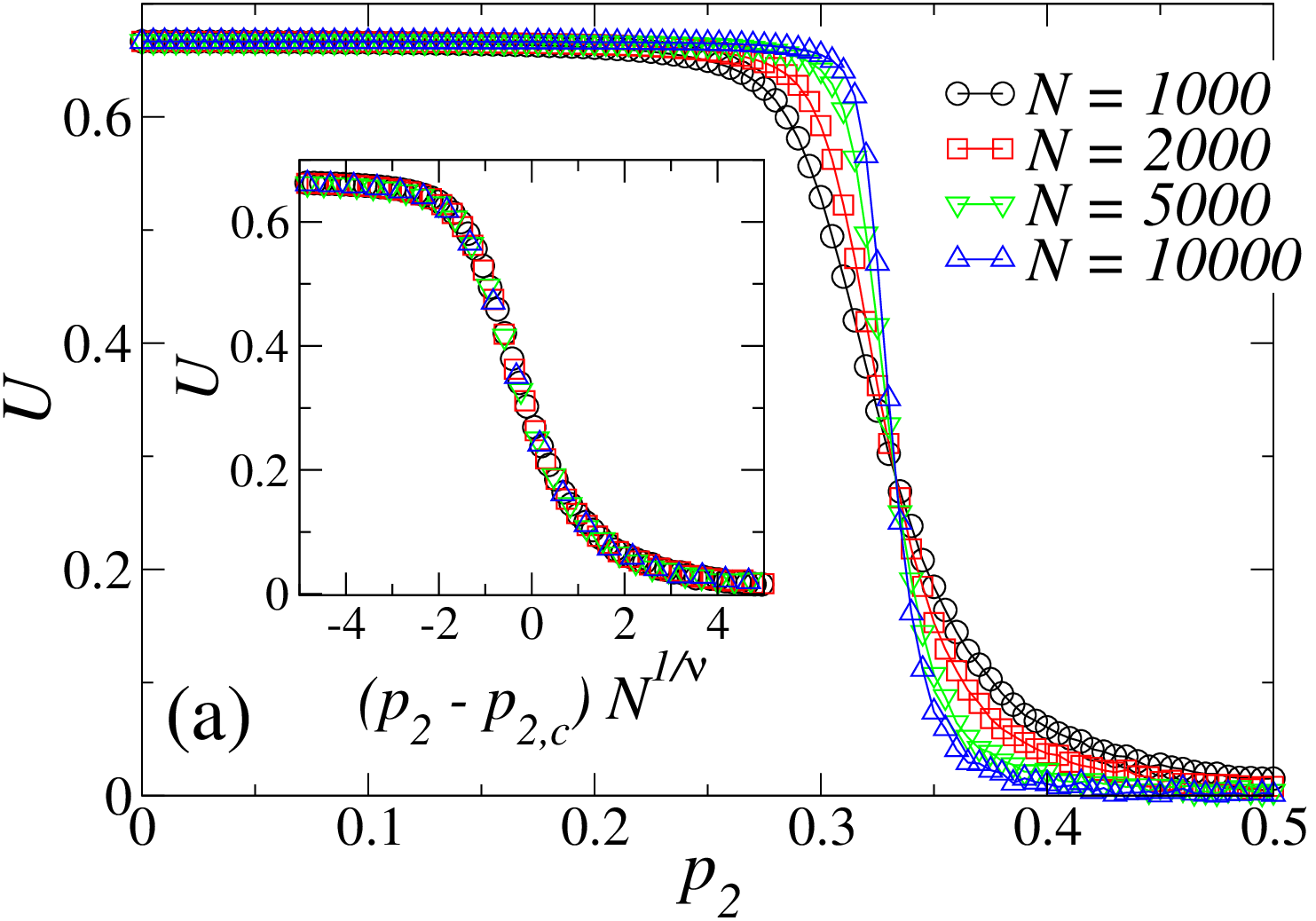}
\hspace{0.3cm}
\includegraphics[width=0.48\textwidth,angle=0]{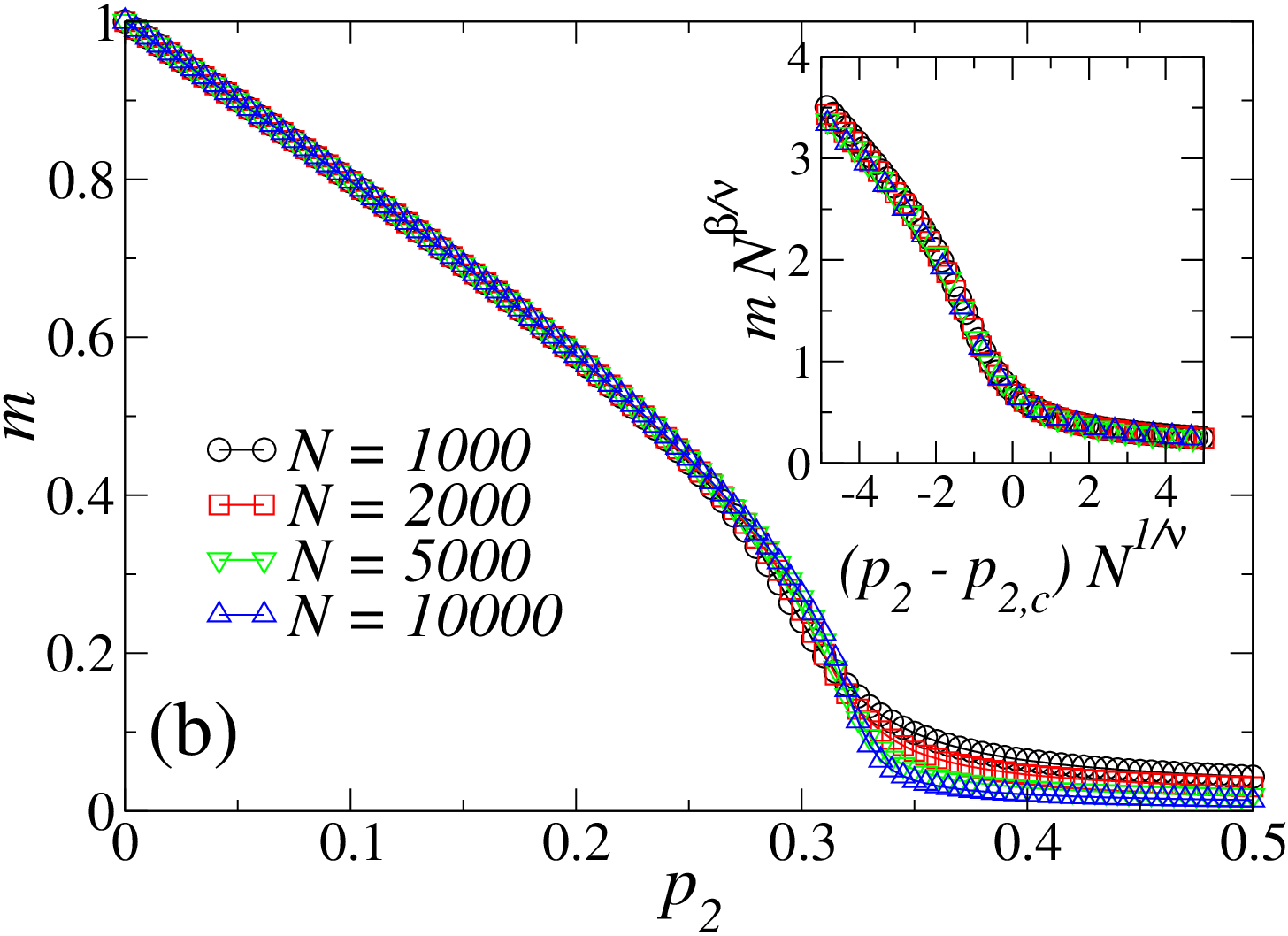}
\\
\vspace{1.0cm}
\includegraphics[width=0.46\textwidth,angle=0]{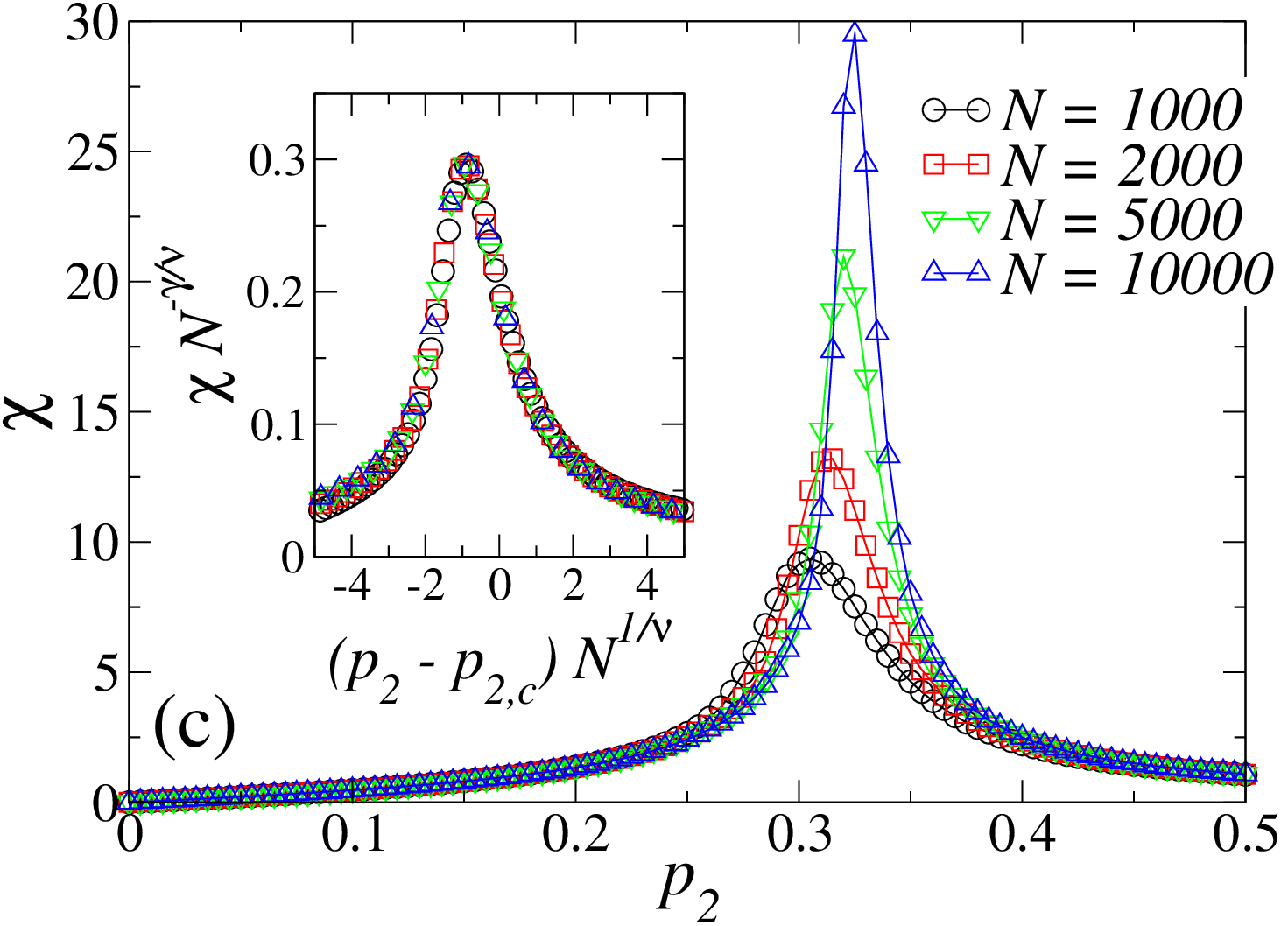}
\end{center}
\caption{(Color online) Numerical results for the mean-field formulation of the Sznajd model with anticonformity for distinct population sizes $N$. Data are for the Binder cumulant $U$ (a), the order parameter $m$ (b) and the susceptibility $\chi$ (c). It is also exhibited in the insets the corresponding scaling plots, based on Eqs. \eqref{eq4}-\eqref{eq7}. We obtained $p_{2,c}\approx 0.33$, $\beta\approx 0.5$, $\gamma\approx 1.0$ and $\nu\approx 2.0$. Data are averaged over $200$ simulations.}
\label{fig1}
\end{figure}

In order to test numerically the analytical results of Ref. \cite{sznajd_anti}, we performed numerical simulations of the model. We exhibit in Fig. \ref{fig1} the numerical results for the Binder cumulant $U$, the order parameter $m$ and the susceptibility $\chi$ as functions of the probability $p_{2}$. We also estimated the critical exponents $\beta$, $\gamma$ and $\nu$. In the insets of Fig. \ref{fig1} it is shown the FSS analysis of data. The critical value $p_{2,c}$ was identified by the crossing of the Binder cumulant curves, as can be seen in the main graphic of Fig. \ref{fig1} (a), and the critical exponents $\beta$, $\gamma$ and $\nu$ were found by the best collapse of data, based on the FSS Eqs. \eqref{eq4}-\eqref{eq7}. We found $p_{2,c}\approx 0.33$, which agrees with the analytical result $p_{2,c}=1/3$ \cite{sznajd_anti}. Concerning the exponents, we also found $\beta\approx 0.5$, $\gamma\approx 1.0$ and $\nu\approx 2.0$ \footnote{Estimates of the error bars are given in Table \ref{Tab1}, obtained from monitoring small variations around the best collapsing pictures.}, which suggests a universality of the order-disorder phase transition. In particular, the numerical estimate of the exponent $\beta$ agrees with Eq. \eqref{eq8}, that predicts $\beta=1/2$. Notice that the exponents $\beta$ and $\gamma$ are typical mean-field exponents, which is not the case for $\nu$. This same discrepancy has already observed in other discrete opinion models \cite{biswas,nuno_celia_victor,nuno_pmco_pre}, and it was associated with a superior critical dimension $d_{c}=4$, that leads to an effective exponent $\nu^{'}=1/2$, obtained from $\nu=d_{c}\,\nu^{'}=2$. In this case, one can say that our model is in the same universality class of the kinetic exchange opinion models with two-agent interactions \cite{biswas,nuno_ijmpc,biswas2}, as well as in the mean-field Ising universality class.


\subsection{Square lattice}

\begin{figure}[t]
\begin{center}
\vspace{3mm}
\includegraphics[width=0.48\textwidth,angle=0]{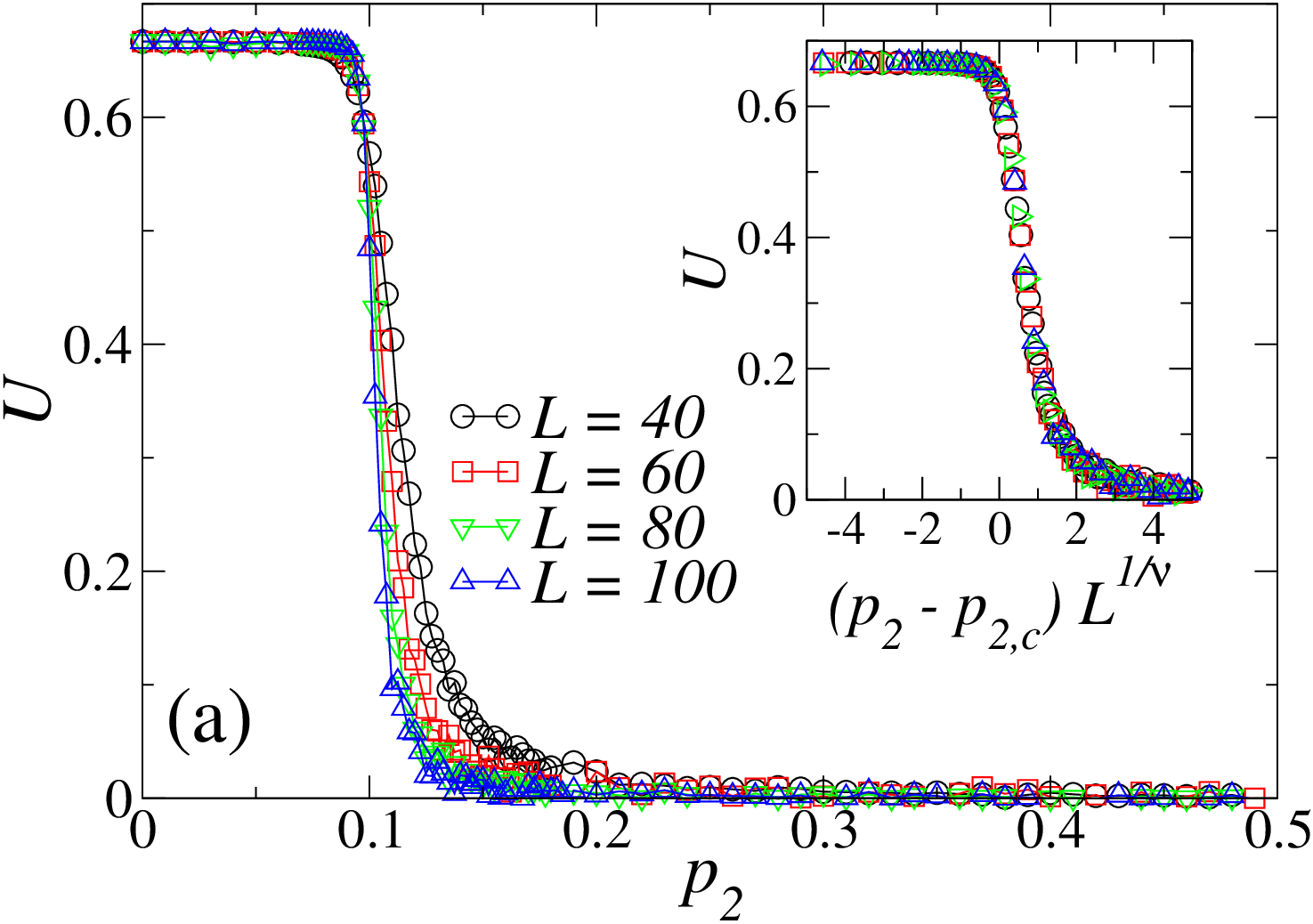}
\hspace{0.3cm}
\includegraphics[width=0.48\textwidth,angle=0]{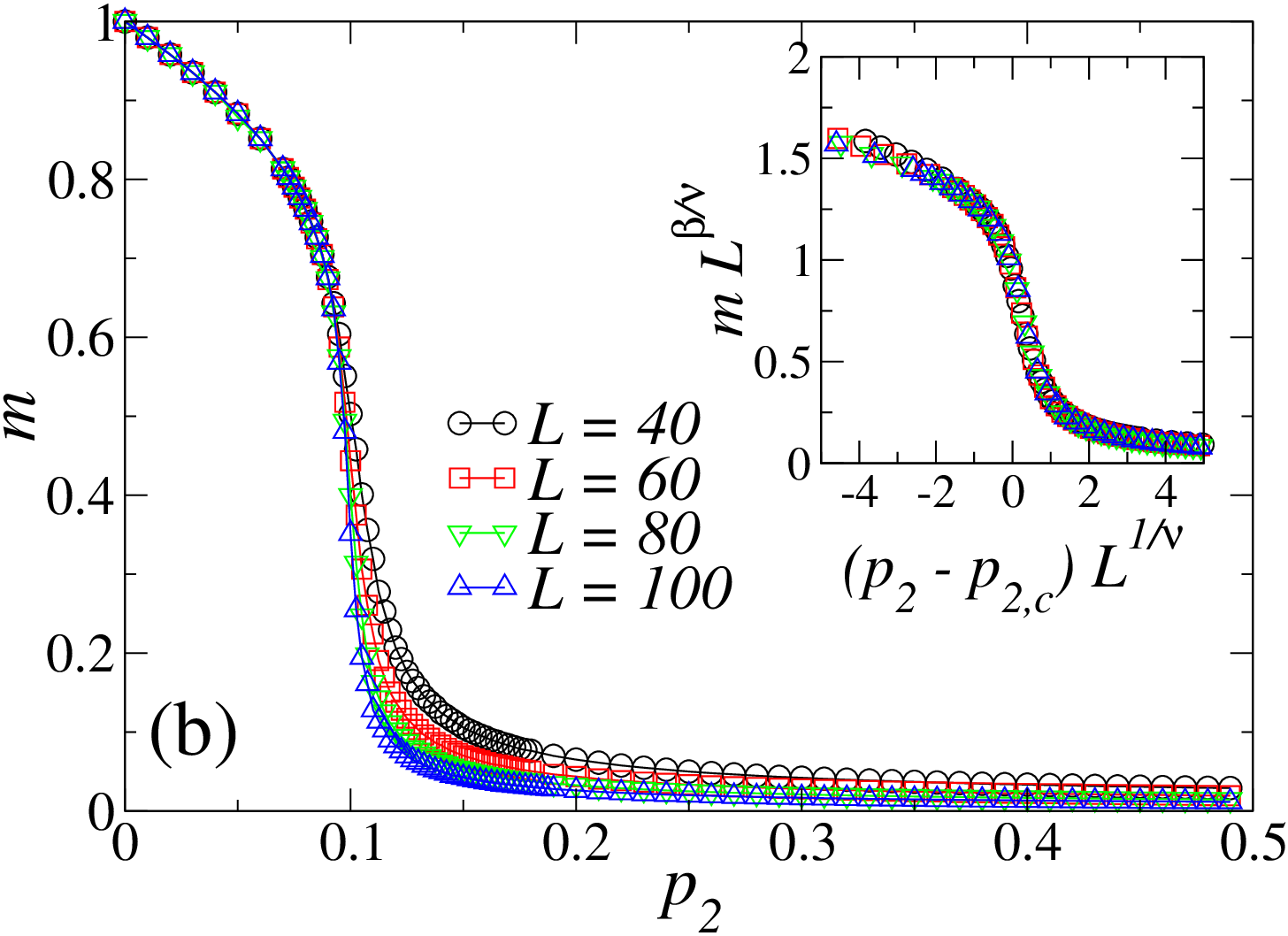}
\\
\vspace{1.0cm}
\includegraphics[width=0.46\textwidth,angle=0]{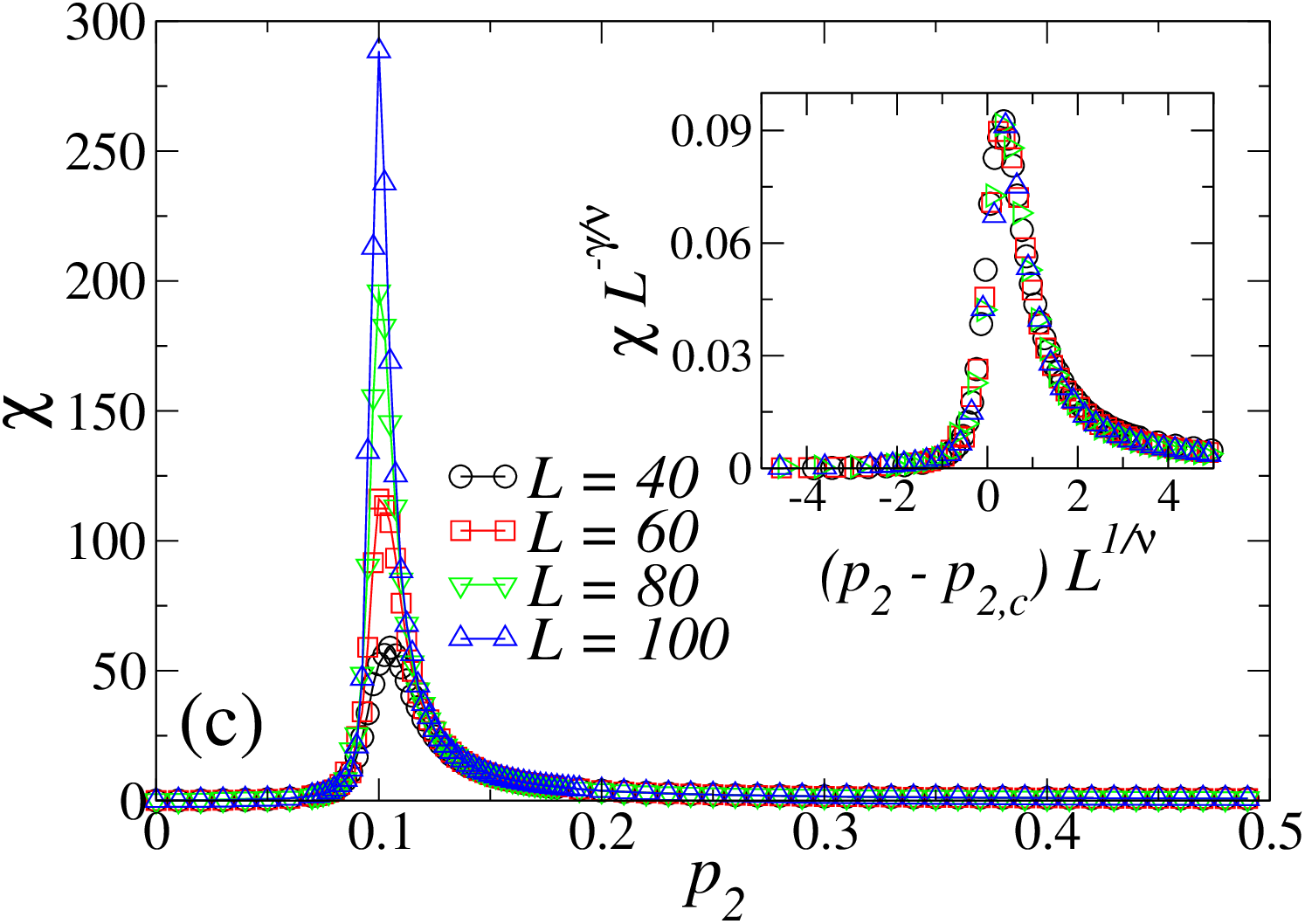}
\end{center}
\caption{(Color online) Numerical results for the two-dimensional formulation of the Sznajd model with anticonformity for distinct linear lattice sizes $L$. Data are for the Binder cumulant $U$ (a), the order parameter $m$ (b) and the susceptibility $\chi$ (c). It is also exhibited in the insets the corresponding scaling plots. We obtained $p_{2,c}\approx 0.096$, $\beta\approx 0.125$, $\gamma\approx 1.75$ and $\nu\approx 1.0$. Results are averaged over $200$, $160$, $120$ and $100$ simulations for $L=40$, $60$, $80$ and $100$, respectively.}  
\label{fig2}
\end{figure}

To test the universality of the model under the presence of a topology, we simulated the dynamics on two-dimensional square lattices of linear size $L$ (total population $N=L^{2}$). In this case, each group of 3 agents $(i,j,k)$ is chosen as follows. First, we choose an agent at random, say $i$. Then, we choose at random two of the four nearest neighbors of $i$ (say $j$ and $k$). We choose this strategy of interaction in order to follow the idea of the original model defined on the fully-connected graph, where 2 agents are (randomly) chosen to influence a third agent (also randomly chosen). We verified that the critical behavior of the model is not affected if we choose more than 2 neighbors, i.e., the model also undergoes a phase transition with the same critical exponents.

Numerical results are exhibited in Fig. \ref{fig2}. Again, the critical value $p_{2,c}$ was identified by the crossing of the Binder cumulant curves, as can be seen in main graphic of Fig. \ref{fig2} (a). We obtained $p_{2,c}\approx 0.096$. The FSS analysis was performed based on Eqs. \eqref{eq4}-\eqref{eq7}, with the change $N \to L$. The exponents were estimated from the best data collapse. We obtained the same critical exponents of the 2d Ising model, i.e., $\beta\approx 0.125$, $\nu\approx 1.0$ and $\gamma\approx 1.75$, suggesting that the model is in fact in the universality class of the Ising model.


\subsection{Cubic lattice}

We also simulated the dynamics on three-dimensional cubic lattices of linear size $L$ (total population $N=L^{3}$). In this case, each group of 3 agents $(i,j,k)$ is chosen as follows. First, we choose an agent at random, say $i$. Then, we choose at random two of the six nearest neighbors of $i$ (say $j$ and $k$).

\begin{figure}[t]
\begin{center}
\vspace{3mm}
\includegraphics[width=0.48\textwidth,angle=0]{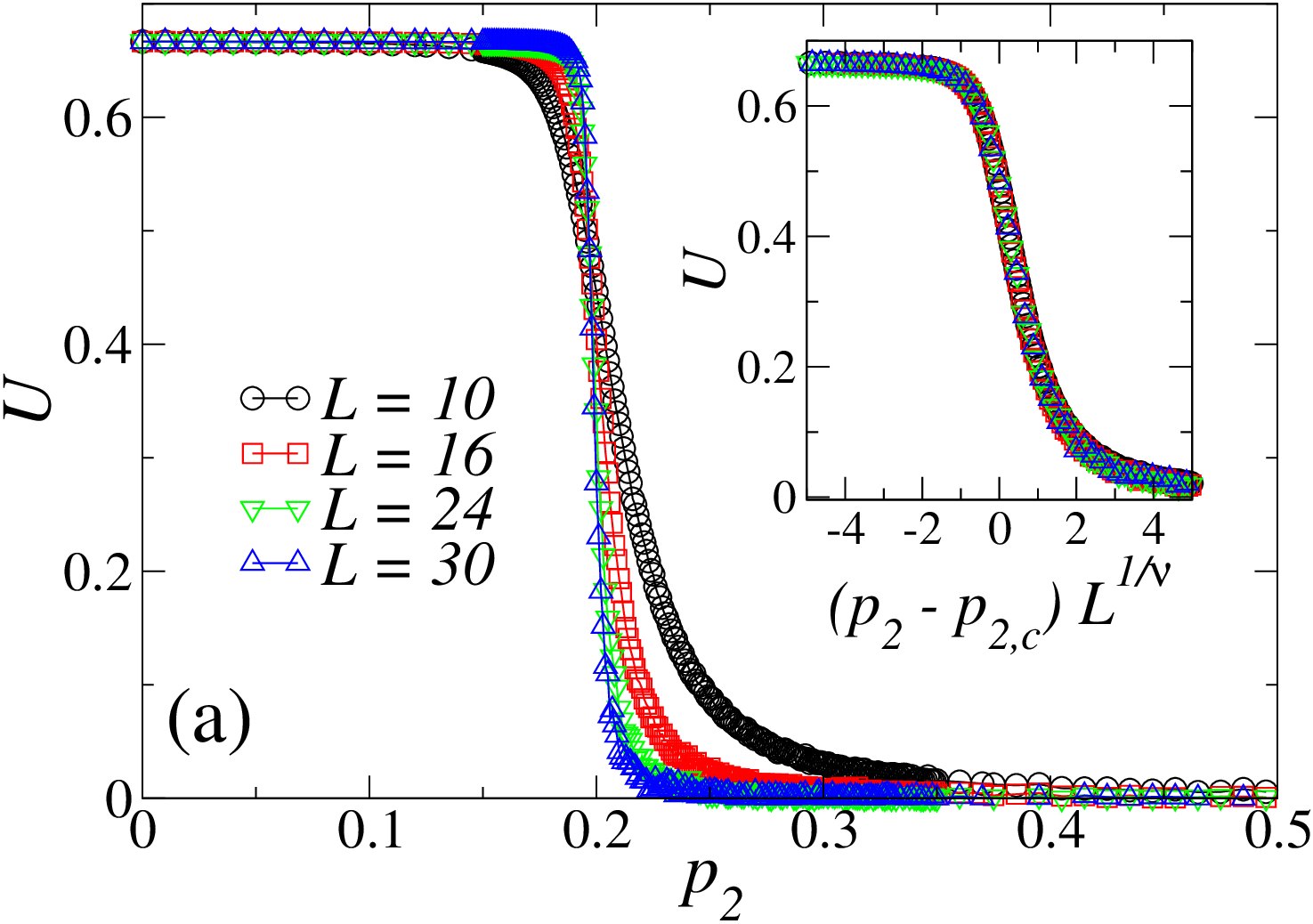}
\hspace{0.3cm}
\includegraphics[width=0.48\textwidth,angle=0]{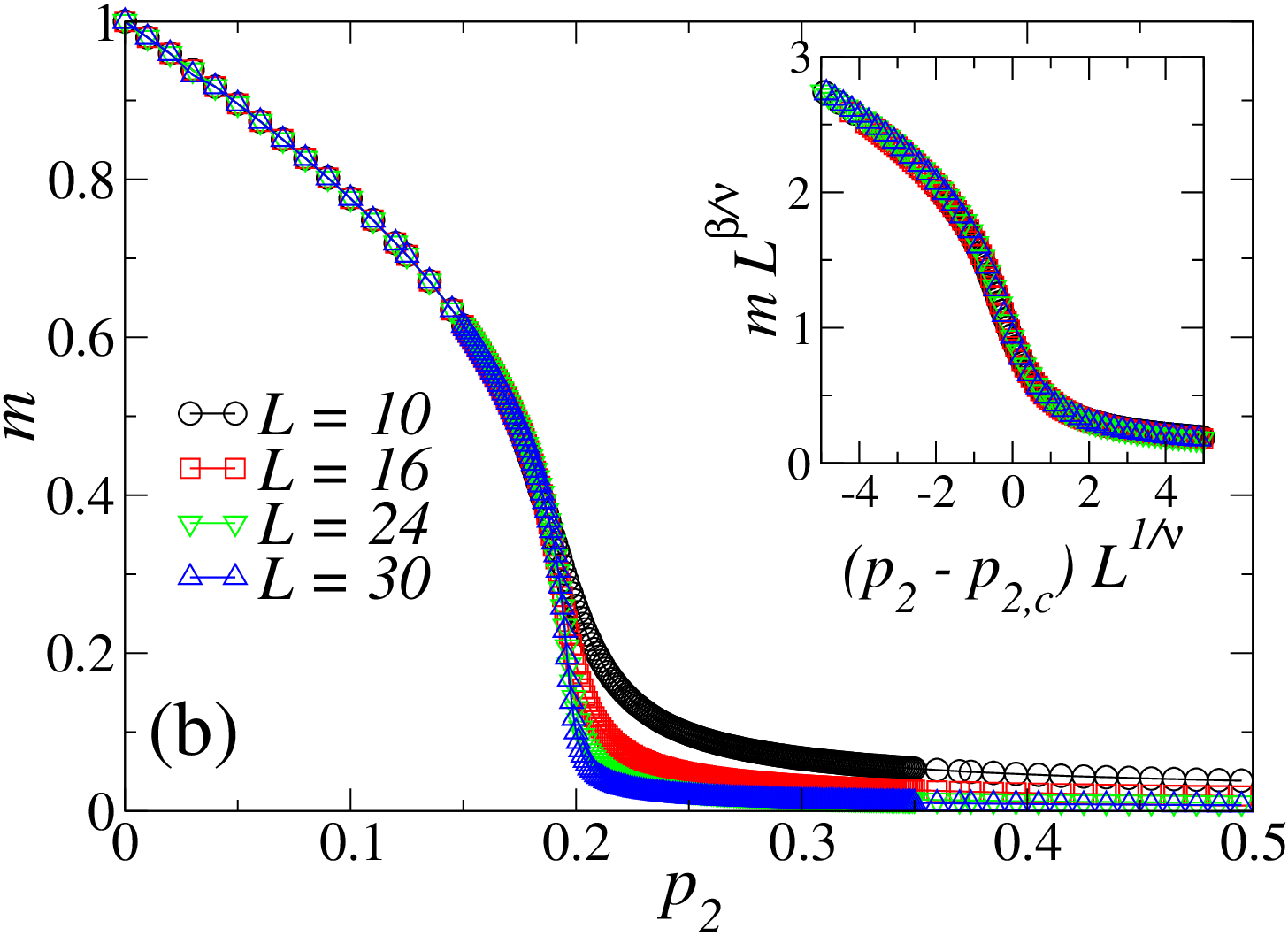}
\\
\vspace{1.0cm}
\includegraphics[width=0.46\textwidth,angle=0]{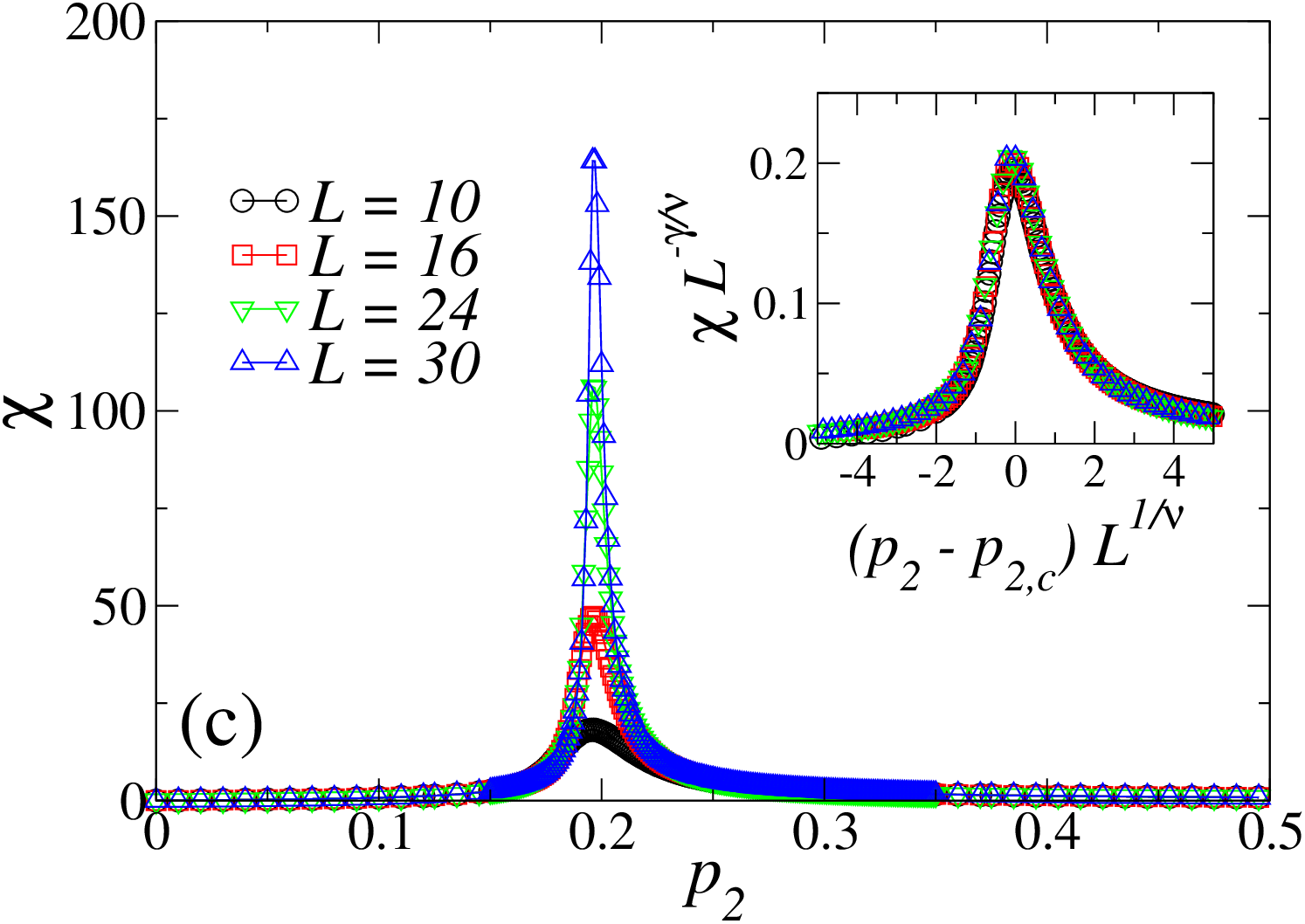}
\end{center}
\caption{(Color online) Numerical results for the three-dimensional formulation of the Sznajd model with anticonformity for distinct linear lattice sizes $L$. Data are for the Binder cumulant $U$ (a), the order parameter $m$ (b) and the susceptibility $\chi$ (c). It is also exhibited in the insets the corresponding scaling plots. We obtained $p_{2,c}\approx 0.197$, $\beta\approx 0.32$, $\gamma\approx 1.24$ and $\nu\approx 0.63$.  Results are averaged over $200$, $175$, $150$ and $125$ simulations for $L=10$, $16$, $24$ and $30$, respectively.} 
\label{fig3}
\end{figure}

\begin{table*}[b]
\begin{center}
\renewcommand\arraystretch{1.3} 
\begin{tabular}{|c|c|c|c|c|}
\hline
$D$ & $p_{2,c}$ & $\beta$ & $\gamma$ & $\nu$     \\ \hline
2 & 0.096\,$\pm$\,0.002 & 0.125\,$\pm$\,0.003 & 1.75\,$\pm$\,0.02 & 1.00\,$\pm$\,0.05  \\ 
3 & 0.197\,$\pm$\,0.002  & 0.32\,$\pm$\,0.01 & 1.23\,$\pm$\,0.02 & 0.62\,$\pm$\,0.02 \\ 
mean field & 0.33\,$\pm$\,0.02 & 0.50\,$\pm$\,0.02 & 1.00\,$\pm$\,0.03 & 2.00\,$\pm$\,0.03 \\ \hline
\end{tabular}%
\end{center}
\caption{The critical probability $p_{2,c}$ that separates the ordered and disordered phases, and the critical exponents $\beta$, $\gamma$ and $\nu$ for distinct lattice dimensions.}
\label{Tab1}
\end{table*}

Numerical results are exhibited in Fig. \ref{fig3}. The estimate of the critical point is $p_{2,c}\approx 0.197$. The FSS analysis was performed based on Eqs. \eqref{eq4}-\eqref{eq7}, with the change $N \to L$. The exponents were estimated from the best data collapse. We obtained the same critical exponents of the 3d Ising model, i.e., $\beta\approx 0.32$, $\nu\approx 0.63$ and $\gamma\approx 1.24$, suggesting that the model is in fact in the universality class of the Ising model.

A summary of all critical values is exhibited in Table \ref{Tab1}.


\section{Final remarks}   

In this work, we have studied a discrete-state opinion model where each agent carries one of two possible opinions. In such case, we have considered Ising-like variables to represent such opinions, i.e., $o_{i}=+1 \,(\uparrow)$ or $o_{i}=-1 \,(\downarrow)$, $i=1,2,...,N$, where $N$ is the population size. The dynamics of the model is based on the Sznajd model, where two mechanisms of social behavior were considered: conformity and anticonformity. Our target was to study the critical behavior of the opinion model under the presence of the mentioned mechanisms.

The dynamics under consideration intends to study the persistence of minority opinions in populations where the anticonformist behavior is present. In the absence of such behavior, the population achieves consensus states with all positive or all negative opinions. Thus, the presence of anticonformity opens the possibility to the coexistence of both opinions in the population, making the Sznajd model more realistic. The phase transition is controled by the parameter $p_{2}$, that is the probability of an agent to act as anticonformist. We verified in all situations analyzed that there is a critical probability $p_{2,c}$ that separates an ordered phase and a disordered phase. The ordered phase represents a collective macroscopic state where both opinions coexist in the population, below the critical probability. On the other hand, above the critical value no majority exists in a system - the so-called stalemate situation. The phase transition is continuous. From the social perspective, particularly interesting is the phenomenon of spontaneous transitions between two states.

Our target was twofold. First, we were interested in study the dynamics on regular lattices in order to observe the emergence of the above-mentioned phase transition when we consider simple structures of social contacts among the individuals. In addition, we were interested in the caracterization of the phase transition, i.e., in the estimation of the critical exponents in distinct dimensions. Thus, we performed computer simulations of the model defined on fully-connected networks, two-dimensional square lattices and three-dimensional cubic lattices. We estimated the critical exponents $\beta$, $\gamma$ and $\nu$ in all cases, through finite-size scaling analysis of data. Our results suggest that the model is in the universality class of the standard Ising model (with no external field) in the respective dimensions.

The presence of special individuals, such as inflexibles or opportunists, has been considered in some opinion models \cite{galam_book,sen_book}. It would be interesting to extend the present model to include the above-mentioned agents and study the impact of their actions on opinion formation. In addition, one can also study the properties of the model in complex networks.


\section*{Acknowledgments}

The authors acknowledge financial support from the Brazilian scientific funding agencies Coordena\c{c}\~ao de Aperfei\c{c}oamento de Pessoal de N\'ivel Superior (CAPES), Conselho Nacional de Desenvolvimento Cient\'ifico e Tecnol\'ogico (CNPq) e Funda\c{c}\~ao Carlos Chagas Filho de Amparo \`a Pesquisa do Estado do Rio de Janeiro (FAPERJ).


\end{document}